\documentclass[prl,10pt,a4paper,twocolumn,noshowpacs]{revtex4}%
\usepackage{amsfonts}
\usepackage{amsmath}
\usepackage{amssymb}
\usepackage{graphicx}%
\begin{document}
\title{Anyons and transmutation of statistics via vacuum induced Berry phase}
\author{Roberto M. Serra$^{1}$, Angelo Carollo$^{1,2}$, Marcelo Fran\c{c}a
Santos$^{1,3}$, and Vlatko Vedral$^{1}$}
\affiliation{$^{1}$ Optics Section, The Blackett Laboratory, Imperial College, London SW7
2BZ, United Kingdom \linebreak$^{2}$ Centre for Quantum Computation, DAMTP,
University of Cambridge, Wilberforce Road, Cambridge CB3 0WA, United Kingdom
\linebreak$^{3}$ Departamento de F\'{\i}sica, Universidade Federal de Minas
Gerais, Belo Horizonte, 30161-970, MG, Brazil}

\begin{abstract}
We show that bosonic fields may present anyonic behavior when interacting with
a fermion in a Jaynes-Cummings-like model. The proposal is accomplished via
the interaction of a two-level system with two quantized modes of a harmonic
oscillator; under suitable conditions, the system acquires a fractional
geometric phase. A crucial role is played by the entanglement of the system
eigenstates, which provides a two-dimensional confinement in the effective
evolution of the system, leading to the anyonic behavior. For a particular
choice of parameters, we show that it is possible to transmute the statistics
of the system continually from fermions to bosons. We also present an
experimental proposal, in an ion-trap setup, in which fractional statistical
features can be generated, controlled, and measured.


\end{abstract}
\maketitle

Anyons are quasiparticles that exhibit fractional quantum statistics which
arises when bosons or fermions are confined in a two-dimensional space
\cite{Wilczek,Canright}. Such excitations play a fundamental role in the
physics of the fractional quantum Hall effect \cite{Wilczek2,Ezawa}. Beside
the theoretical applications, anyonic systems are promising for the
implementation of fault-tolerant quantum computing \cite{kitaev}.

In this paper, we show how typical bosons, specifically eigenstates of
harmonic oscillators (number states), behave as fractional spin particles,
so-called anyons, when coupled to a two-level system (a fermion) through a
Jaynes-Cummings-like interaction.

We can interpret the origin of such an exotic behavior as a consequence of the
entanglement capability of the interaction Hamiltonian (the entangled nature
of the Hamiltonian eigenstates). When the coupled system is subjected to an
adiabatic evolution that mixes the orthogonal modes of the harmonic
oscillator, the entanglement provided by the Hamiltonian can be regarded as a
two-dimensional constraint on one of the two subsystems, allowing for
\emph{bosonic excitations }of fields (photons or phonons) to behave like
anyons. This fractional behavior is clearly manifested in the geometric phase
\cite{Berry,Shapere} acquired by any eigenstate of the coupled system under a
cyclic adiabatic evolution, and it is possible to relate the geometrical phase
with the statistical factor \cite{Berry2}.

The bosonic field behaves, in an analogous way, as a spin-1 particle when
there is no interaction (in the sense that the geometric phase acquired during
the evolution is the same as that acquired by a spin-1 particle in a magnetic
field), to a spin-1/4 particle when the two systems are maximally entangled,
going through all the intermediate spins depending on the specific degree of
entanglement between the two subsystems. More specifically, it is possible to
simulate anyons with $\frac{m}{2}$,$\frac{m}{3}$,$\frac{m}{4}$,... ($m=1,$
$2,...$) statistics or even to transmute continually the statistics of the
system from Fermi to Bose.

We also present a proposal, employing trapped ions, in which fractional
statistical phases can be generated, manipulated, and tested. This proposal,
in addition to its experimental interest, provides a clearer physical
framework to interpret the presented theoretical results.

Initially, we consider an effective two-level (fermionic) system coupled
nonlinearly to two quantized bosonic fields, through the so-called $m$-quantum
Jaynes-Cummings model \cite{Vogel}. In the rotating-wave approximation, this
Hamiltonian is given by ($\hbar=1$)%
\begin{equation}
H=\nu a^{\dagger}a+\nu b^{\dagger}b+\frac{\omega}{2}\sigma_{z}+\lambda
_{m}\left[  \sigma_{+}\left(  a\right)  ^{m}+\sigma_{-}\left(  a^{\dagger
}\right)  ^{m}\right]  ,\label{eq1}%
\end{equation}
where $\sigma_{z}=\left\vert \uparrow\right\rangle \left\langle \uparrow
\right\vert -\left\vert \downarrow\right\rangle \left\langle \downarrow
\right\vert ,$ $\sigma_{+}=\left\vert \uparrow\right\rangle \left\langle
\downarrow\right\vert ,$ and $\sigma_{-}=\left\vert \downarrow\right\rangle
\left\langle \uparrow\right\vert $ are the usual Pauli pseudospin operators
($\left\vert \uparrow\right\rangle $ and $\left\vert \downarrow\right\rangle $
are the excited and ground states of the two-level system, respectively),
$\lambda_{m}$ is the effective $m$ nonlinear coupling constant, $a^{\dagger}$
($a$) and $b^{\dagger}$ ($b$) are the creation (annihilation) operators of the
bosonic fields with frequency $\nu$.

In the interaction picture and in a rotating frame (through the unitary
transformation $\exp\left[  i\Delta_{m}\sigma_{z}t/2\right]  $), $H$ can be
rewritten as%
\begin{equation}
\mathbf{H}=\frac{\Delta_{m}}{2}\sigma_{z}+\lambda_{m}\left[  \sigma_{+}\left(
a\right)  ^{m}+\sigma_{-}\left(  a^{\dagger}\right)  ^{m}\right]  ,\label{eq2}%
\end{equation}
with $\Delta_{m}=\omega-m\nu$ being the effective detuning between the bosonic
and fermionic systems. Field mode $a$ is orthogonal to mode $b$ and,
initially, the two-level system interacts only with the former one.

The eigenstates of the joint system, associated to the eigenvalues $\pm
\Lambda$ ($\Lambda=\left[  \left(  \Delta_{m}\right)  ^{2}/4+\left(
\lambda_{m}\right)  ^{2}\left(  n+m\right)  !/n!\right]  ^{1/2}$), are
described by
\begin{equation}
\left\vert \Psi_{n,n^{\prime}}^{\pm}\right\rangle =\left(  C_{\uparrow
}\left\vert \uparrow\right\rangle \left\vert n\right\rangle _{a}\pm
C_{\downarrow}\left\vert \downarrow\right\rangle \left\vert n+m\right\rangle
_{a}\right)  \left\vert n^{\prime}\right\rangle _{b},\label{eq3}%
\end{equation}
where%
\[
C_{\uparrow}=\frac{\Lambda+\left.  \Delta_{m}\right/  2}{\sqrt{2}\sqrt
{\Lambda^{2}+\Lambda\left.  \Delta_{m}\right/  2}},\quad C_{\downarrow}%
=\frac{\lambda_{m}\sqrt{\left(  n+m\right)  !/n!}}{\sqrt{2}\sqrt{\Lambda
^{2}+\Lambda\left.  \Delta_{m}\right/  2}},
\]
and $\left\vert n\right\rangle _{l}$ ($l=a,b$) is the Fock state of the
bosonic field $l$.

Let us consider the case in which the Hamiltonian described by Eq. (\ref{eq1})
is slowly transformed into a parametrized one, $\mathbf{H}(\theta
,\phi)=U(\theta,\phi)\mathbf{H}U^{\dag}(\theta,\phi)$. This transformation is
achieved by externally driving the interacting systems, and if it is done
adiabatically, i.e., slowly compared to the typical time scales of the
problem, the eigenstates of the initial Hamiltonian follow the transformation,
as defined in the adiabatic theorem. In the end of a closed cycle (i.e.,
$\theta$ and $\phi$ are cyclically varied and then brought back to their
original value), each eigenstate goes back to the original one, except for a
phase factor of geometrical nature $e^{i\gamma_{n,n^{\prime}}}$\textbf{
}\cite{Berry,Shapere}.

We are interested in the parametric transformation of Hamiltonian (\ref{eq1})
obtained through the unitary operation
\begin{equation}
U\left(  \theta,\phi\right)  =\exp(-i\phi J_{z})\exp(-i\theta J_{y}%
),\label{eq4}%
\end{equation}
where $J_{z}=\left(  a^{\dagger}a-b^{\dagger}b\right)  /2$ and $J_{y}=i\left(
ab^{\dagger}-a^{\dagger}b\right)  /2$ are the Schwinger angular momentum
operators. In this case, for each value of the parameters $\theta$ and $\phi$,
this transformation results in the two-level system interacting with a linear
combination of the two bosonic fields, $a$ and $b$ \cite{Fuentes,Angelo}.

A suitably slow variation of the parameters $\theta$ and $\phi$ results in the
adiabatic evolution of the eigenstates $\left\vert \Psi_{n,n^{\prime}}^{\pm
}\right\rangle $ of the initial Hamiltonian $\mathbf{H}$ in Eq. (\ref{eq2}) .
When $\theta$ and $\phi$ are eventually brought back to their initial values,
$\left\vert \Psi_{n,n^{\prime}}^{\pm}\right\rangle $ acquires a Berry phase,
which is given by%
\begin{align}
\gamma_{n,n^{\prime}} &  =i%
{\displaystyle\int\nolimits_{c}}
d\phi d\theta\left\langle \Psi_{n,n^{\prime}}^{\pm}\right\vert U^{\dagger
}\left(  \theta,\phi\right)  \nabla_{\theta,\phi}U\left(  \theta,\phi\right)
\left\vert \Psi_{n,n^{\prime}}^{\pm}\right\rangle \nonumber\\
&  =\Omega\left\langle \Psi_{n,n^{\prime}}^{\pm}\right\vert J_{z}\left\vert
\Psi_{n,n^{\prime}}^{\pm}\right\rangle \nonumber\\
&  =\frac{\Omega}{2}\left(  n-n^{\prime}+\left(  \frac{m}{2}\right)
\frac{\left(  \lambda_{m}\right)  ^{2}\left(  n+m\right)  !/n!}{\left(
\Lambda^{2}+\Lambda\frac{\Delta_{m}}{2}\right)  }\right)  ,\label{phase-off}%
\end{align}
where $\Omega$ is the solid angle subtended by the cyclic path in the
Poincar\'{e}'s sphere. Both eigenstates $\left\vert \Psi_{n,n^{\prime}}%
^{+}\right\rangle $ and $\left\vert \Psi_{n,n^{\prime}}^{-}\right\rangle $
acquire the same geometrical phase.

We are interested in the case of zero initial\ excitation in the bosonic
fields ($n=n^{\prime}=0$), and the most interesting scenario is achieved when
$\Delta_{m}=0$, i.e., the resonant interaction. In this regime, the harmonic
oscillator and the two-level particle are exchanging $m$ excitations and the
geometric phase reduces to $\gamma_{0,0}=\frac{m}{4}\Omega$. Notice that, in
this expression, each order of the non-linear interaction in (\ref{eq1}) is
contributing with a factor of $1/4$ to the total geometric phase. The $1/4$
factor originates in the fermion-boson interaction as shown in~\cite{Fuentes},
while the $m$ factor is typical for a collective behavior. In this sense, the
Hamiltonian~(\ref{eq1}) can be regarded as a multibody interaction, between
the two-level system and $m$-bosonic particles.

The closed path described by the transformation (\ref{eq4}) can be regarded as
two consecutive exchanges of bosonic excitations in the two possible modes,
which here play the role of two different spatial configurations of one
anyonic particle. For example, initially the two-level system is exchanging
$m$ excitations with mode $a$. Then, for $\theta=\pi$, the exchange involves
only mode $b$, and finally for $\theta=2\pi$, a complete cycle is done and the
interaction is back to mode $a$. This is analogous to the case of one electron
orbiting around a magnetic flux tube in the original Wilczek work
\cite{Wilczek}. The analogy with the classical rotation in real space is
clearer in the trapped ions example presented later, where modes $a$ and $b$
can be chosen as nothing but spatial vibration modes of the trap (see  Ref.
\cite{BEC} for the implementation of fractional dynamics in the context of
Bose-Einstein condensation).

In analogy with the physical two-dimensional space, we exploit in our system
the parametric space of the Hamiltonian. When the parameters of the
Hamiltonian $\mathbf{H}(\theta,\phi)$ are changed and eventually returned to
their original configuration, the wave function remains the same, except for
the phase factor $e^{i\alpha2\pi}$, where $\alpha=\gamma_{0,0}/2\pi$ is called
the statistical factor. In a 3-dimensional space, the rotation group satisfies
a peculiar non-Abelian algebra, which allows only for the Bose ($\alpha=0$) or
Fermi statistics ($\alpha=1$). On the other hand, if we restrict ourselves to
2-dimensional rotations, the corresponding group can generate a broader class
of (braid) statistics \cite{Forte}. It turns out, in fact, that the
impossibility of using rotations in a third dimension reduces the number of
constraints on the symmetry of the wave function, allowing for fractional
values of $\alpha$.

We can interpret the eigenstate of the system described by Eq. (\ref{eq3}) (an
entangled state of the two-level system and the field) as a kind of
two-dimensional confinement, in the sense that only the rotation $J_{z}$
contributes for the Berry connection in Eq. (\ref{phase-off}), introducing the
fractional behavior. To deepen our understanding of the role played by the
entanglement of the eigenstates of the system (\ref{eq3}), we will analyze the
statistical factor $\alpha$ in different regimes, $\Delta_{m}\gg\lambda_{m}$,
$\Delta_{m}\sim\lambda_{m}$ and $\Delta_{m}=0$. In general, the statistical
factor is given by
\begin{equation}
\alpha=\frac{m}{4}\frac{\Omega}{2\pi}\frac{\left(  \lambda_{m}\right)  ^{2}%
m!}{\left(  \Lambda^{2}+\Lambda\frac{\Delta_{m}}{2}\right)  }, \label{factor}%
\end{equation}
and presents a dependence on the order $m$ of the nonlinear interaction, on
solid angle $\Omega$, and also on the detuning $\Delta_{m}$.

Let us consider the far off-resonance case, $\Delta_{m}\gg\lambda_{m}$, which
means that the harmonic oscillators and the two-level system are not
exchanging energy, i.e., the eigenstate of the Hamiltonian is completely
separable. This is the most trivial case, where $\alpha=0$ and the excitations
of the harmonic oscillators behave as bosons, as we should expect for a
noninteracting system~\cite{Fuentes}. On the other hand, in the resonant
case\ ($\Delta_{m}=0$) the eigenstate of Hamiltonian \ (\ref{eq2}) is
maximally entangled and the statistical factor $\alpha=\frac{m}{4}\frac
{\Omega}{2\pi}$. Therefore, the fractional features of the system's geometric
phase depend crucially on the entangled form of the system's eigenstate. In
Fig 1(a), we show the ratio $\gamma_{0,0}\left(  \frac{m}{4}\frac{\Omega}%
{2\pi}\right)  ^{-1}$ as a function of the detuning $\Delta_{m}$, and in Fig.
1(b) we show the linear entropy of the reduced two-level system in state
(\ref{eq3}) $S_{f}=1-Tr_{f}\left[  \left(  Tr_{b}\rho\right)  ^{2}\right]  $
(where $f$ and $b$ stand for fermionic and bosonic variables, respectively,
and $\rho=\left\vert \Psi_{0,0}^{\pm}\right\rangle \left\langle \Psi
_{0,0}^{\pm}\right\vert $) as a function of the detuning $\Delta_{m}$.
\ Comparing these two figures, we can see that, when the detuning increases,
the amount of entanglement decreases and consequently the fractional features
of the system also decrease, until they disappear completely for separable eigenstates.%

\begin{figure}
[t]
\begin{center}
\includegraphics[
height=4.9865in,
width=3.4701in
]%
{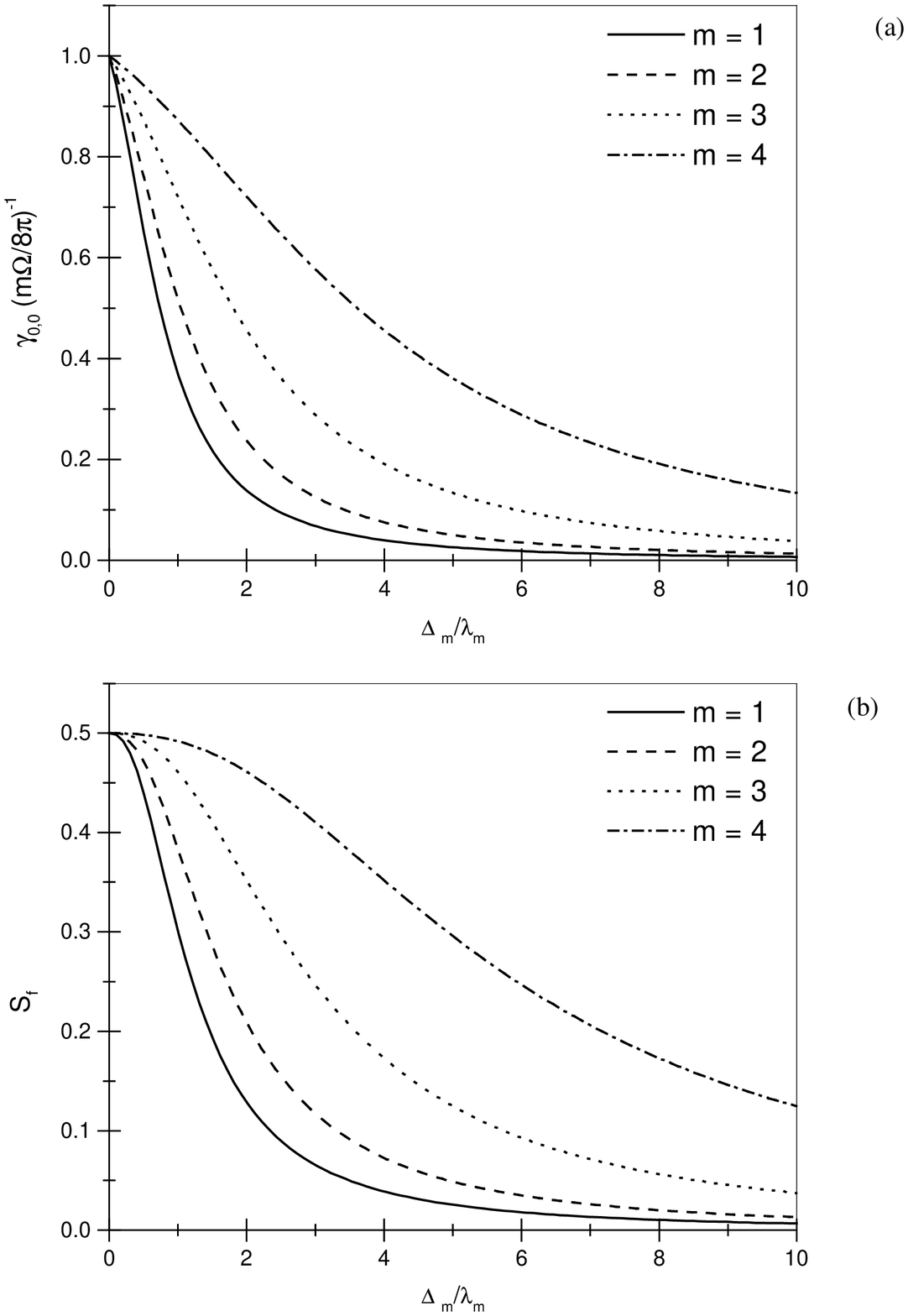}%
\caption{(a) The fractional \ part of the geometrical phase $\left\{
\gamma_{0,0}\left(  \frac{m}{4}\Omega\right)  ^{-1}\right\}  $ and (b) the
linear entropy of the reduced two-level system in the state (\ref{eq3}) as
function of the effective detuning $\Delta_{m}$.}%
\end{center}
\end{figure}

It is interesting to note that, if we vary $\theta$, $\phi$ in a way that the
cyclic path in Poincar\'{e}'s sphere encloses the whole sphere ($\Omega=4\pi
$), the statistical factor, for $\Delta_{m}=0$, turns out to be $\alpha
=\frac{m}{2}$, if we choose a path that encloses $2/3$ of the sphere,
$\alpha=\frac{m}{3}$; and so on.\ In this way, we can simulate anyons with
$\frac{m}{2}$,$\frac{m}{3}$,$\frac{m}{4}$,... statistics. Let us consider the
interesting particular situation (for $\Delta_{m}\neq0$) in which we choose
$m=2$ and vary the parameters $\theta$, $\phi$ in order to obtain $\Omega
=4\pi$. In that way, the statistical factor depends only on the detuning
$\Delta_{2}$. In the resonant case ($\Delta_{2}=0$), we have $\alpha=1$ and
the system obeys the Fermi statistics. On the other hand, in the far detuned
case ($\Delta_{2}\gg1$) the statistical factor tends to zero ($\alpha
\rightarrow0$) and the system obeys the Bose statistics. In this particular
situation, varying the detuning $\Delta_{2}$ we can transmute continually the
statistics of the system from Fermi to Bose, going through the anyonic
statistics. The statistical factor $\alpha$ depends on $\Delta_{2}$ with the
same pattern showed in Fig. 1(a) for $m=2$ (dashed line). We note that for the
case $m=1$ the value $\alpha=1$ cannot be reached. In this latter case, the
statistical factor can vary continually from $0$ to $1/2$ depending on the
detuning $\Delta_{1}$ for the fixed solid angle $\Omega=4\pi$.

The scenario presented here for the simulation of one anyon statistics can be
generalized for two anyons. Since the system composed by a fermion interacting
with two bosonic fields behaves like one anyon, the generalization to two
anyons can be achieved by adding two more bosonic fields ($c,d$). Let us
suppose that the dynamics of the system is governed by the parametrized
Hamiltonian (in the interaction picture) $\widetilde{\mathbf{H}}(\theta
,\phi)=\lambda_{m}\widetilde{U}(\theta,\phi)\left[  \sigma_{+}\left(
a\right)  ^{m}\left(  c\right)  ^{m}+\sigma_{-}\left(  a^{\dagger}\right)
^{m}\left(  c^{\dagger}\right)  ^{m}\right]  \widetilde{U}^{\dag}(\theta
,\phi)$, under the unitary transformation $\widetilde{U}(\theta,\phi
)=\exp\left[  -i\phi\left(  J_{z}^{ab}+J_{z}^{cd}\right)  \right]  \exp\left[
-i\theta\left(  J_{y}^{ab}+J_{y}^{cd}\right)  \right]  $ (where $J_{z}^{lk}$
and $J_{y}^{lk}$ are the Schwinger angular momentum operators for modes $l$
and $k$). Under cyclic and suitably slow variation of parameters $\theta$ and
$\phi$, the eigenstates of this new Hamiltonian, given by $\left\vert
\widetilde{\Psi}_{0}^{\pm}\right\rangle =\left(  \left\vert \uparrow
\right\rangle \left\vert 0\right\rangle _{a}\left\vert 0\right\rangle _{c}%
\pm\left\vert \downarrow\right\rangle \left\vert m\right\rangle _{a}\left\vert
m\right\rangle _{c}\right)  \left\vert 0\right\rangle _{b}\left\vert
0\right\rangle _{d}\left/  \sqrt{2}\right.  $, evolve adiabatically. If
$\theta$ and $\phi$ are eventually brought back to their initial values,
$\left\vert \widetilde{\Psi}_{0}^{\pm}\right\rangle $ acquires the Berry phase
$\widetilde{\gamma}_{0}=\frac{m}{2}\Omega$, which corresponds to twice the
phase obtained in the previous case. Consequently, for the two-anyons case we
have the statistical factor $\alpha=\frac{m}{2}\frac{\Omega}{2\pi}$, which is
exactly twice as much as the one described in Eq. (\ref{factor}) for
$\Delta_{m}=0$.

Finally, we discuss how to implement, in a physical context, Hamiltonian
(\ref{eq1})\ and the unitary transformation (\ref{eq4}). To this end we
consider one single ion in a two-dimensional harmonic electromagnetic trap in
the $x$,$y$ plane, with degenerate frequency $\nu$. The ion has two effective
electronic states, $\left\vert \uparrow\right\rangle $ and $\left\vert
\downarrow\right\rangle $, separated by the frequency $\omega_{0}$ and coupled
by the interaction with an effective laser plane wave propagating initially in
the $x$ direction, with frequency $\omega_{L}$ and wavevector $\overrightarrow
{k}_{L}=\left(  \omega_{L}/c\right)  \overrightarrow{x}$. In this
configuration, only the ionic motion along the $x$ axis will be modified and
the Hamiltonian of this system is given by \cite{Nist}
\begin{equation}
H=\nu a^{\dagger}a+\nu b^{\dagger}b+\frac{\omega_{0}}{2}\sigma_{z}+g\left[
\sigma_{+}e^{i\overrightarrow{k}_{L}\cdot\overrightarrow{r}-i\omega
_{L}t+i\varphi_{L}}+\text{\textrm{H.c.}}\right]  ,
\end{equation}
where $\overrightarrow{r}$ is the vibration direction of the ion, $g$ the
effective coupling constant for transition $\left\vert \uparrow\right\rangle $
$\longleftrightarrow$ $\left\vert \downarrow\right\rangle $, and $\varphi_{L}$
is the phase of the laser \cite{Aus}. The effective laser beam is tuned to the
$m$-th red vibrational sideband of the ion, i.e., it is detuned by
$\delta=\omega_{0}-\omega_{L}=m\nu$ from the $\left\vert \uparrow\right\rangle
$ $\longleftrightarrow$ $\left\vert \downarrow\right\rangle $ transition. In
the interaction picture and in the frame rotating at the effective laser
frequency $\omega_{L}$, the Hamiltonian of this system can be written as
\cite{Nist,Matos}%
\begin{equation}
\mathbf{H}=f_{m}\left(  \eta,a^{\dagger}a\right)  e^{i\varphi_{L}}\sigma
_{+}\left(  a\right)  ^{m}+\text{\textrm{H.c.},}%
\end{equation}
where $f_{m}\left(  \eta,a^{\dagger}a\right)  =\frac{1}{2}ge^{-\eta^{2}%
/2}\left[  \sum_{l=0}^{\infty}\frac{\left(  i\eta\right)  ^{2l+m}}{l!\left(
l+m\right)  !}\left(  a^{\dagger}\right)  ^{l}\left(  a\right)  ^{l}%
+\mathrm{H.c.}\right]  $ is the effective coupling, $\eta=\sqrt{\left.
\left(  \overrightarrow{k}_{L}\cdot\overrightarrow{r}\right)  ^{2}\right/
2\mathcal{M}\nu}$ is the Lamb-Dicke parameter, and $\mathcal{M}$ is the ion
mass. When we assume the so-called Lamb-Dicke regime $\eta\ll1$, we have
$f_{m}\left(  \eta,a^{\dagger}a\right)  \sim\frac{1}{2\left(  m!\right)
}g\left(  i\eta\right)  ^{m}e^{-\eta^{2}/2}\equiv\lambda_{m}$. Different
values of the nonlinear interaction order $m$ can be reached by the choice of
the $m$-th red vibrational sideband of the ion. In this way we can implement
Hamiltonian (\ref{eq1}). If we consider the effective pumping laser beam tuned
not exactly on the $m$-th vibrational sideband of the ion, we can obtain the
off-resonance case of\ Eq. (\ref{eq1}) with a small detuning limited by the
frequencies of the neighboring side bands ($\Delta_{m}<<\nu$). In this
context, the statistical transmutation of the system can be investigated
varying the solid angle $\Omega$ that depends on the path choice for
parameters $\theta$, $\phi$. The unitary transformation (\ref{eq4}) can be
implemented by ($i$) the physical rotation, in the $x$,$y$ plane, of the
propagation direction of the effective laser field, in a way that
$\overrightarrow{k}_{L}=\left(  \omega_{L}/c\right)  \left[  \cos\left(
\left.  \theta\right/  2\right)  \overrightarrow{x}+\sin\left(  \left.
\theta\right/  2\right)  \overrightarrow{y}\right]  ,$ together with ($ii$) a
suitable variation on phase $\varphi_{L}$ (note that $\varphi_{L}=2\varphi$).
Therefore, by the control of the propagation direction of the laser and its
phase, it is possible to implement physically the unitary transformation
(\ref{eq4}).

The fractional phase acquired by the ion due to unitary transformation
(\ref{eq4}) can be measured by a Ramsey-type interferometer similar to the one
suggested in~\cite{Angelo}. Let us start with the assumption that the
vibrational modes of the ion in both directions $x$,$y$ are cooled to their
ground states, and its electronic internal levels\ are prepared in a
superposition $\left.  \left(  \left\vert \uparrow\right\rangle +\left\vert
\downarrow\right\rangle \right)  \left\vert 0\right\rangle _{a}\left\vert
0\right\rangle _{b}\right/  \sqrt{2}$. This superposition can be generated by
a carrier-type laser pulse \cite{Nist}, which corresponds to the choice $m=0$
(the laser is tuned in the $\left\vert \uparrow\right\rangle $
$\longleftrightarrow$ $\left\vert \downarrow\right\rangle $ transition
frequency, $\omega_{0}=\omega_{L}$), and does not affect the vibrational modes
of the ion. The next step consists of turning on a laser tuned to the $m$-th
red vibrational sideband and initially aligned in the $x$ direction with
reference phase $\varphi_{L}=0$. This laser interacts with the ion for a time
$\tau\gg\left.  1\right/  \lambda_{m}$, during which its propagation direction
is rotated and $\varphi_{L}$ is cyclically changed. The variation rate of
these parameters must be much smaller than the effective coupling $\lambda
_{m}$ to allow the adiabatic regime assumed in our approach. When the laser is
turned off, the system has evolved to state $\left.  \left(  e^{i\gamma_{0,0}%
}\left\vert \uparrow\right\rangle +\left\vert \downarrow\right\rangle \right)
\left\vert 0\right\rangle _{a}\left\vert 0\right\rangle _{b}\right/  \sqrt{2}%
$; here we have used the fact that both eigenstates $\left\vert \Psi
_{n,n^{\prime}}^{+}\right\rangle $ and $\left\vert \Psi_{n,n^{\prime}}%
^{-}\right\rangle $ acquire the same geometrical phase $\gamma_{0,0}$ and
$\left\vert \uparrow\right\rangle \left\vert 0\right\rangle _{a}\left\vert
0\right\rangle _{b}=\left.  \left(  \left\vert \Psi_{0,0}^{+}\right\rangle
+\left\vert \Psi_{0,0}^{-}\right\rangle \right)  \right/  \sqrt{2}$. It is
clear that the variation of parameters $\theta$ and $\varphi_{L}$ must be
designed such that the end of its cyclic path coincides with the end of the
$j$-th Rabi cycle performed in the time $\tau=\left.  2\pi j\right/
\lambda_{m}$ ($j\gg1$). Finally, we turn on again a carrier type laser pulse
($m=0$) with phase $\varphi_{L}=\pi/2$ during time $t=\left.  \pi\right/
(2g)$. After this rotation we perform a fluorescence measurement of the
electronic states of the ion \cite{Nist}, with probability $P_{\downarrow
}=\left(  1-\cos\left(  \gamma_{0,0}\right)  \right)  /2$ to measure level
$\left\vert \downarrow\right\rangle $. In this way, we can measure the
fractional phase introduced by the above procedure.

In this paper, we have presented a method to simulate the dynamics of anyons
via the Jaynes-Cummings model. We have shown how to simulate anyons with
$\frac{m}{2}$,$\frac{m}{3}$,$\frac{m}{4}$,... statistics or even how to
transmute continually the statistics of the system from Fermi to Bose, going
through the anyonic statistics. Such fractional features depend crucially on
the entangled form of the system's eigenstate which works as a two-dimensional
confinement. We also provide a proposal for the physical implementation of the
above ideas in the context of trapped ions phenomena. It introduces a novel
possibility for the investigation of fractional statistics in a
well-controlled way.

\begin{acknowledgments}
This research was supported by the Engineering and Physical Sciences Research
Council, the European Commission, and Elsag-spa company. M.F.S. and R.M.S.
acknowledge the support of CNPq.
\end{acknowledgments}

\end{document}